\documentstyle[12pt]{article}

\begin{document}

\newpage
\noindent
\hfill{\small $forarxiv/np1dbox.tex$}

\vspace{1cm}
        
\centerline {\bf GROUND STATE OF A SYSTEM OF N HARD CORE }

\smallskip
\centerline {\bf QUANTUM PARTICLES IN 1-D BOX}
\vspace{.3cm}

\centerline {\bf Yatendra S. Jain}

\smallskip

\centerline {Department of Physics, North-Eastern Hill University,}

\centerline {Shillong-793 022, Meghalaya, India}

\bigskip
\noindent
\begin{abstract}
The ground state of a system of $N$ impenetrable hard core
quantum particles in a 1-D box is analyzed by using a new
scheme applied recently to study a similar system of two
such particles {\it [Centl. Eur. J. Phys., 2(4), 709 (2004)]}.
Accordingly, each particle of the system behaves like an
independent entity represented by a {\it macro-orbital}, -a
kind of pair waveform identical to that of a pair of
particles moving with ($q$, $-q$) momenta at their
{\it center of mass} which may have any momentum $K$ in the
laboratory frame.  It concludes: (i) $<A\delta{(x)}> = 0$, (ii)
$<x> \ge \lambda/2$ and (iii) $q \ge q_o (= \pi/d)$ (with $d
= L/N$ being the average nearest neighbor distance), {\it etc.}
While all bosons in their ground state have $q = q_o$ and $K = 0$,
fermions have $q= q_o$ with different $K$ ranging between $0$
and $K = K_F$ (the Fermi wave vector).  Independent of their
bosonic or fermionic nature, all particles in the ground state
define a close packed arrangement of their equal size wave
packets representing an ordered state in phase ($\phi-$)space
with $\Delta\phi = 2n\pi$ (with $n$ = 1,2,3, ...), $<x> =
\lambda/2 = d$, and $q = q_o$.   As such our approach uses
greatly simplified mathematical formulation and renders a
visibly clear picture of the low energy states of the systems
and its results supplement earlier studies in providing their
complete understanding.

\end{abstract}



{\it email}: ysjain@email.com

\bigskip

\noindent
{\bf 1. Introduction}

\bigskip
Wave mechanics of a 1-D system of $N$ identical particles
interacting through {\it hard core} (HC) potential of zero
or non-zero range has been a subject of great importance for
the last seven decades [1-21].  The development of the subject
has been facilitated by the fact (demonstrated by first few
studies by Tonks [1], Girardeau [2,3], and Lieb and Liniger
[4]) that the problem is exactly solvable.  Identical
conclusion has been made later [22, 23] for a similar system
of particles interacting through inverse square potential.
These developments have been elegantly reviewed by Popov [24],
Korepin, {\it et al} [25], Mattis [26] and most recently by
sutherland [27].  As evident from the number of such studies
[28-44], interest in the subject is renewed recently after the
use of {\it Bose Einstein condensation} (BEC) in dilute gases
in experimental realization of many body systems [45-54] that
can be described by 1-D Hamiltonian with $\delta -$potential
interaction [55, 56].  The scheme of theoretical analysis of
these systems uses Bethe ansatz [57] for $N$ body wave
function with bosonic/fermionic symmetry (as the case
demands) and suitable (periodic/cyclic) boundary conditions.
Proposed initially by Lieb and Liniger [4], properties of
such a system can be determined theoretically by introducing
a dimensionless parameter, $\gamma = mg_{1D}/n\hbar^2$ (with
$m$,-the mass of a particle, $n$, -the number density of
particles and $g_{1D}$, -the 1-D coupling constant related
to 3-D $s-$wave scattering length $a$).  One identifies
Tonks-Girardeau (TG) regime of strong coupling (large
$g_{1D}$) and low density ($n$) by $\gamma \approx \infty$,
Bogoliuobov regime [56] (or decoherent regime [53]) of
nearly free bosons by $\gamma \approx 0$, and Lieb and
Liniger (LL) regime of weak coupling (low $g_{1D}$) and
high density by intermediate values ({\it viz.}, $\infty >
\gamma > 0$).  One also defines Gross-Pitaevskii (GP)
regime [53] by $1 > \gamma > 0$.  Apparently, the behavior
of a system of given $g_{1D}$, can be studied in its
different regimes simply by changing $n$ from $0$ to
$\infty$.  One of the important conclusions of these
studies reveals that the physics of a bosonic system
characterized by $\gamma \approx \infty$ resembles with that
of free fermions.  

\bigskip
In this paper, we analyze the {\it ground state} (G-state)
of a 1-D system of $N$ HC particles by using a slightly
different scheme.  The excitations of the system would be
investigated in our forthcoming article.  Our scheme uses
the wave mechanics of a pair of HC particles ({\it
identified as the basic unit of the system}) as its basis;
it has been investigated in our recent paper [58] for two
HC particles trapped in a 1-D box [58].  Our
approach renders exact solutions. In view of the
equivalence of impenetrable $\delta-$repulsion with HC
interaction of non-zero range [{\it i.e.}, $V_{HC}(x)$,
defined by $V_{HC}(x < \sigma) = \infty$ and $V_{HC}(x \ge
\sigma) = 0$ with $\sigma$ being the HC diameter of a
particle] demonstrated by Huang [59] and physically argued
in [58], one may find that our results for $\delta
$-repulsion can be applied to particles of any $\sigma$,
particularly, in the low energy states of $\lambda/2 \ge
\sigma$.

\bigskip
Our results differ from those of extensively used scheme(s)
of others [24-27], particularly, for the low energy states
where the well known consequences of wave particle duality,
{\it viz.}, inter-particle phase correlation, zero-point
repulsion, quantum size, {\it etc.} (defined and discussed
in Section-3 for their better understanding), become
important.  Our scheme provides a mathematically simple,
unified, and clear picture of 1-D systems.  As such one may
find that the present study supplements earlier studies in
providing a better understanding of our 1-D systems and its
basic results resemble with those of : (i) a single particle
in 1-D box [60, 61], two HC particles in a similar box ([58],
and (iii) long awaited microscopic theory of a 3-D system of
interacting bosons [62-64] which explains the properties of
liquid $^4He$ (a well known representative of such systems)
with unmatched accuracy, simplicity and clarity.  This
demonstrates the accuracy and usefulness of our scheme.  As
reported briefly in [65, 66], our scheme provides a
theoretical framework that unifies the physics of widely
different systems (including low dimensional systems) of
interacting bosons and fermions, atomic nucleus, newly
discovered BEC state, {\it etc.} and helps in revealing
the basic foundations of the microscopic theory of
superconductivity.
 
\bigskip
The paper has been arranged as follows.  While the
important aspects of our N-body system are described in
Section-2, the important inferences of our study of two
HC particles in 1-D box, which serves the basis of this
work, are summed up in Section-3.  This is followed by a
detailed analysis of the ground state of the system in
Section-4 and concluding remarks in Section-5.

\bigskip
\noindent
{\bf 2. Important Aspects of N-Body System}

\bigskip
The hamiltonian of a 1-D system of $N$ particles interacting
through a two body impenetrable $\delta$-potential can be
expressed as
$$H(N) = -{\hbar^2\over{2m}}\sum_i^N{\partial^2\over\partial
^2_{{\rm x}_{i}}} +
\sum_{i>j}A\delta{(x_{ij})} \eqno(1)$$
\noindent
where $A$, representing the strength of Dirac delta function
($\delta{(x_{ij})}$), is such that $A \to \infty$ with $x_{ij}
\to 0$, while other notations have usual meaning.  Possible
interactions such as spin-spin interaction are not included in
Eqn. 1 by presuming that these can be treated perturbatively.

\bigskip
Analysing the basic nature of the possible dynamics of two
particles (say P1 and P2) in the system, we note that: (i) P1
and P2 encounter $\delta$-potential only when they suffer a
collision, (ii) in a two body collision, they simply exchange
their momenta, ${\rm k}_1$ and ${\rm k}_2$, while in a many
body collision, where they also collide simultaneously with
other particle(s), they could be identified to jump from
their state of ${\rm k}_1$ and ${\rm k}_2$ (or their relative
momentum $k = {\rm k}_2-{\rm k}_2$ and {\it center of mass}
(CM) momentum $K = {\rm k}_2+{\rm k}_2$) to that of new
momenta ${\rm k'}_1$ and ${\rm k'}_2$ (or $k'$ and $K'$),
and (iii) between two collisions each of them has free
particle motion.  Evidently, a state of two particles even
in the presence of other particles of our system can be
identified characteristically with the state of only two HC
particles in 1-D space.  In other words a pair of particles
forms the basic unit of our system and this agrees with the
fact that particles in the system interact through a pair
potential.

\bigskip
\noindent
{\bf 3. Dynamics of two HC particles (P1 and P2)}

\bigskip
We note that the dynamics of P1 and P2 can be described by
the Schr\"{o}dinger equation for $H(2)$ [Eqn. 1 with $N=2$]
expressed in the CM coordinate system as
$$\left(-{{\hbar}^2\over {4m}}{\partial^2\over \partial^2_X}
-{{\hbar}^2\over {m}}{\partial^2\over \partial^2_x}
+ A\delta{(x)}\right){\Psi}(x,X) =
E{\Psi}(x,X), \eqno(2)$$
\noindent
with 
$${\Psi}(x,X) = {\psi}_k(x)\exp{[i(KX)]}    \eqno(3)$$

\noindent
where ${\psi}_k(x)$ and $\exp{[i(KX)]}$, respectively,
define the relative and CM motions of P1 and P2.  We have
$$x = {\rm x}_2 - {\rm x}_1 \quad {\rm and} \quad
k = {\rm k}_2 - {\rm k}_1 = 2q, \eqno(4) $$
$$ X = ({\rm x}_1 + {\rm x}_2)/2 \quad {\rm and} \quad
K = {\rm k}_1 + {\rm k}_2 \eqno(5) $$

\noindent
with $x$ and $k$, respectively, describing the relative
position and relative momentum of P1 and P2, while $X$
and $K$, similarly, defining the position and momentum of
their CM.  Without loss of generality, one may also define
$${\rm k}_1 = -q + \frac{K}{2} \quad {\rm and} \quad
{\rm k}_2 = q + \frac{K}{2}.  \eqno(6)$$

\noindent
It is evident that ${\psi}_k(x)$ is a solution of 
$$\left(-{{\hbar}^2\over {m}}{\partial^2\over
\partial^2_x} + A\delta{(x)}\right){\psi}_k(x) =
E_k{\psi}_k(x)    \eqno(7)$$

\noindent
where $E_k = E - {\hbar}^2K^2/4m$.  With a view to find
${\psi}_k(x)$, we note that P1 and P2 experience zero
interaction at $x \not = 0$ where each of them can be
described by separate plane waves such as $u_{{\rm k}_i}
({\rm x}_i, t) = \exp{(i{\rm k}_i{\rm x}_i)}
\exp{(-iE_it/\hbar)}$ (assumed to have unit normalization),
however, for the possible superposition of these waves,
their quantum state is better expressed by
$$\Psi{({\rm x}_1, {\rm x}_2, t})^{\pm} = 1/\sqrt{2}.
[u_{{\rm k}_1}({\rm x}_1,t)u_{{\rm k}_2}({\rm x}_2,t)
\pm u_{{\rm k}_2}({\rm x}_1,t)u_{{\rm k}_1}({\rm x}_2,t)].
\eqno(8)$$

\noindent
Since the state function of two impenetrable HC particles
must vanish at ${\rm x}_1 = {\rm x}_2$, $\Psi{({\rm x}_1,
{\rm x}_2, t})^+$ (having $+ve$ symmetry for their exchange)
does not represent the desired function, while
$\Psi{({\rm x}_1, {\rm x}_2, t})^-$ of $-ve$ symmetry has no
such problem.   We addressed this problem in our recent
study [58] of the wave mechanics of two impenetrable HC
particles in 1-D box and used another method to obtain the
right wave function of $+ve$ symmetry.  Evidently, using our
results of [58], we express a state of two HC
fermions/bosons (in the CM coordinates) by
$${\zeta}(x,X,t)^{\pm} = \zeta_k(x,t)^{\pm}\exp{[i(KX)]}
\exp{[-i(E_k + E_K)t/\hbar]}    \eqno(9)$$

\noindent
with 
$$\zeta_k(x,t)^- = \sqrt{2}\sin{(kx/2)}\exp{[-iE_kt/\hbar]}
\eqno(10)$$

\noindent
of fermionic symmetry and 
$${\zeta_k}(x,t)^+ =
\sqrt{2}\sin{(|kx|/2)}\exp{[-iE_kt/\hbar]} \eqno(11)$$

\noindent
of bosonic symmetry.  Note that ${\zeta}(x,X,t)^-$ is a form
of $\Psi{({\rm x}_1, {\rm x}_2, t})^-$ (Eqn. 8) expressed in
CM coordinates.  Although, two impenetrable HC particles in
1-D are expected to retain the order of their locations
({\it e.g.} $0 < {\rm x}_1 < {\rm x}_2 < L$ for two particles
in a 1-D box with infinite potential walls located at
${\rm x} = 0$ and ${\rm x} = L$) but Eqns. 9-11 remain valid
for describing the state of P1 and P2 since the two particles
during their collision do exchange ${\rm k}_1$ and ${\rm k}_2$
which is equivalent to exchanging sides because the particles
are identical.  In fact in a state of wave mechanical
superposition of two identical particles, one has no means to
ascertain whether the particles exchanged their momenta or
their positions and we may, justifiably, analyze Eqns. 9-11
to find important aspects of the wave mechanics two HC
particles as follows.

\bigskip
\noindent
(i) {\it SMW state :} ${\zeta}(x,X,t)^{\pm}$ represents a
kind of {\it standing matter wave} (SMW) which modulates the
probability, $|{\zeta}(x,X,t)^{\pm}|^2 = |{\zeta_k}(x,t)
^{\pm}|^2$, of finding the two HC bosons (or fermions) at
their relative phase position $\phi = kx$ in $\phi-$space.
However, the equality, $|{\zeta_k}(x,t)^-|^2$ =
$|{\zeta_k}(x,t)^+|^2$, renders an important fact that the
relative configuration of P1 and P2 in ${\zeta_k}(x,t)^{\pm}$
states is independent of their fermionic or bosonic symmetry.
Although, $t-$dependent terms in Eqns. 9-11 have their own
importance and they help in having a better understanding of
the SMW nature of ${\zeta}(x,X,t)^{\pm}$, nevertheless these
can be dropped for their no impact on our impending analysis.

\bigskip
\noindent
(ii) {\it Symmetry of relative motion :} The relative motion
of P1 and P2 maintains a center of symmetry at their CM and
this implies that
$${\rm x}_{CM}(1) = - {\rm x}_{CM}(2) = x/2 \quad {\rm and}
\quad {\rm k}_{CM}(1) = - {\rm k}_{CM}(2) = q \eqno(12)$$

\noindent
where ${\rm x}_{CM}$ and ${\rm k}_{CM}$, respectively,
represent the position and momentum of a particle with
reference to the CM of P1 and P2.  Evidently, two particles
in ${\zeta_k}(x,t)^{\pm}$ state have equal and opposite
momenta $(q, -q)$ with respect to their CM whose motion
in the laboratory frame can be defined by a plane wave
($\exp(iKX)$) of momentum $K$ and this agrees with what
we learn from Eqn. 6.

\bigskip
\noindent
(iii) {\it MS and SS states :}  We note that
${\zeta}(x,X,t)^{\pm}$ resulting from the superposition
of two plane waves of ${\rm k}_1$ and ${\rm k}_2$ could be
identified to represent the {\it mutual superposition} (MS)
of P1 and P2 because ${\zeta}(x,X,t)^{\pm}$ is, basically,
an eigenstate of the energy operators of the relative
and CM motions of two particles rather than of individual
particle.  However, it could also be identified as a state
of the {\it self superposition} (SS) of individual particle
(P1 or P2) by identifying ${\zeta}(x,X,t)^{\pm}$ as a
{\it macro-orbital} ({\it cf.} Point-vii below) on the
basis of the following analysis.  To understand the
meaning of self superposition one may track down the
motions of P1 and P2 separately and find that each
of them (say P1), after its collision with P2, has
superposition of its pre-collision plane wave $u_{{\rm k}_1}
({\rm x}_1,t)$ with its post-collision plane wave
$u_{{\rm k}'_1}({\rm x}_1,t)$ (with its new momentum
${\rm k}'_1 = {\rm k}_2$ because during the collision P1
exchanges its momentum with P2 and {\it vice versa});
similar observation applies to P2.  As such
${\zeta_k}(x,t)^{\pm}$ part of our SS state
is not different from the state of a particle
trapped in 1-D box.  However,
since P1 and P2 are identical particles we have no
means to identify whether the two particles have
their mutual superposition or a self superposition of
individual particle.  Consequently, ${\zeta}(x,X,t)^{\pm}$
could be used identically to represent either the MS state
of P1 and P2 or the SS state of either particle.  It
may be mentioned that two plane waves of two particles can,
in principle, have their superposition independent of
their relative position $x$ and relative momentum $k$,
however, as shown by experiments the wave nature of
particles seem to influence the behavior
of a system of particles only when their $\lambda =
2\pi/q$ compares with their $x$ which implies that an
effective superposition of two particles leading to their
MS/SS state exists only in low energy state with 
$\lambda \approx x$.

\bigskip
\noindent
(iv) {\it Characteristics of relative motion : }
${\zeta_k}(x,t)^{\pm}$ has a series of antinodes of size
$\lambda/2$ between different nodes at $x = \pm n\lambda/2$
($n = 0,1,2,3,
...$) where $\lambda = 2\pi/q$ represents the de Broglie
wave length related to the relative motion of P1 and P2.
Evidently, two particles can be confined to the {\it shortest
possible space} of $\lambda$ size without disturbing their
${\zeta_k}(x,t)^{\pm}$ state (say, by placing two
impenetrable potential walls at the nodal points,
$x = \pm \lambda/2$ and for such a confinement we 
have [58] $<x> = I'/I = \lambda/2$ with integrals $I' =
<{\zeta_k}(x,t)^{\pm}|x|{\zeta_k}(x,t)^{\pm}>$ and
$I = <{\zeta_k}(x,t)^{\pm}|{\zeta_k}(x,t)^{\pm}>$ performed
between $x=0$ (the minimum possible $x$) to $x = \lambda$
(the maximum possible $x$).  Evaluating similar results
for $<\phi>$ (as shown in [58]), we find that a state
two HC particles in free space is characterized by
$$<x> \ge \lambda/2,  \quad\quad
{\rm and} \quad \quad<\phi> \ge 2\pi.    \eqno(13)$$

\noindent
This implies that any experiment, that keeps track of
the relative dynamics of two HC particles, would find that
P1 and P2 never reach closer than  $<x>_o = \lambda/2$
(shortest possible $<x>$) and $<\phi>_o = 2\pi$ (shortest
possible $<\phi>$).  Similarly, Eqns. 12 and 13 reveal
that the position of P1 and P2 as seen from their CM can
not be closer than $<{\rm x}_{CM}(1)>_o = - \lambda/4$ and
$<{\rm x}_{CM}(2)>_o = \lambda/4$ or {\it vice versa}.
Finally, we also find that
$$<A\delta{(x)}> = |\zeta_k(x)^{\pm}|^2_{x=0} = 0,
\eqno(14)$$

\noindent
which has been analyzed for its general validity in
[58] (see Appendix-A of arXiv.org/quant-ph/0603233) 
which clearly shows that $<A\delta{(x)}> = 0$ is
valid for all physically relevant situations rendering 
$$<H(2)> = (\hbar^2/4m)(K^2 + k^2) =
(\hbar^2/2m)({\rm k}_1^2 + {\rm k}_2^2).
\eqno(15)$$

\bigskip
\noindent
(v) {\it Quantum size of a particle :}  In view of Eqn. 13,
it is evident that two particles by themselve do not assume
a state of $<x> < \lambda/2$ (with $\lambda= 2\pi/q$) which
indicates that each of them exclusively occupies $\lambda/2$
space; being identical particles with equal $|q|$, they are
expected to share $\lambda$ space equally.  Hence as used
in this paper we identify $\lambda/2$ as the effective size
of a particle particularly for low energy states of
$q \le \pi/\sigma$ and name it as {\it quantum size}.  For a
better understanding of this meaning, one may consider the
wave associated with P1 as a probe to scale the size of P2 or
{\it vice versa} and apply the principle of image resolution
to find that P1 (or P2) would be able to resolve the $\sigma$
size of P2 (or P1) only if $\lambda/2 \le \sigma$ ({\it i.e.}
$q \ge \pi/\sigma$) implying that in such a case they see
each other as particles of size $\sigma$.  However. in case
$\lambda/2 > \sigma$, they can not resolve $\sigma$ and would
see each other as the objects of size $\lambda/2$ limited by
their capacity to resolve.  As such the effective size of P1
and P2 is a $q-$dependent quantity for $\lambda/2 > \sigma$
(or $q < \pi/\sigma$) and $q-$independent quantity for
$\lambda/2 \le \sigma$ (or $q \ge \pi/\sigma$).  Evidently,
the use of quantum size for $\lambda/2$ is justified to
distinguish the two situations.  This meaning seems to
qualitatively agree with the meaning of $``${\it quantum
spread of a particle}$''$ as used by Huang [59].  However,
on quantitative scale it seems to represent the minimum
possible quantum spread ({\it limited by the uncertainty
principle}) of a particle of momentum $q$ but it, evidently,
differs from the meaning of $``$quantum size$''$ as used in
relation to the {\it quantum size effects} on the properties
of thin films and small clusters of atoms [67].  The fact
that a particle in its ground state in a box has a spread
of $\lambda/2 = d$ ({\it the size of the box}) also
indicates that the particle exclusively occupies $\lambda/2$
space because any attempt to decrease this space size (say
by reducing $d$) would push the particle to have new 
momentum/energy but once again the space occupied would be 
a new $\lambda/2$.

\bigskip
\noindent
(vi) {\it Zero point repulsion :}  Since two HC particles do
not assume a state of $<x> < \lambda/2$ ({\it cf.}, Eqn. 13),
one would obviously like to identify the responsible force.
To this effect we note that $<x> < \lambda/2$ represents an
overlap two particles of effective size $\lambda/2$ and this
implies that particles in this state have higher energy in
comparison to the state of $<x> \ge \lambda/2$.  Evidently, the
two particles in the state of $<x> < \lambda/2$ experience a
kind of mutual repulsion (or {\it the zero point repulsion})
unless they assumes a state of $<x> \ge \lambda/2$.
Thus $\lambda/2$ represents the effective size of a HC particle
of $q < \pi/\sigma$ and the range of zero-point repulsion.  

\bigskip
\noindent
(vii) {\it Macro-orbital representation of a particle : }
Although, two particles in $\zeta{(x,X)}^{\pm}$ states can
be identified to have inter-particle phase correlation
($g(\phi ) = |\zeta{(x,X)}^{\pm}|^2 = |\sin{(\phi/2)}|^2$)
which can keep them locked at $<\phi> = 2n\pi$ $(n = 1,2, ...)$
in $\phi -$ space, nevertheless they do not form a bound pair in
$x-$space because they either experience mutual repulsion when
$<x> < \lambda/2$ or no force when $<x> \ge \lambda/2$.
Evidently, each of them in $\zeta{(x,X)}^{\pm}$ state represents
an independent entity described by a separate pair waveform, say,
${\xi}(x_{(i)}, X_{(i)}) \equiv \zeta(x,X)$; the subscript $(i)$
refers to $i-$th particle.  To distinguish
${\xi}(x_{(i)}, X_{(i)})$ from  $\zeta(x,X)$, we propose to call
the former a macro-orbital [58] and obtain the same by replacing
$x$, $X$, $k$ and $K$ in $\zeta(x,X))$ by $x_{(i)}$,
$X_{(i)}$, $k_{(i)}$ and $K_{(i)}$, respectively.  This renders   
$${\xi}(x_{(i)}, X_{(i)}) = B\zeta_{q_{(i)}}(x_{(i)})
\exp{[i(K_{(i)}X_{(i)})]} \eqno(15)$$

\noindent
where B is the normalization constant and $\zeta_{q_{(i)}}
(x_{(i)})$ is that part of a macro-orbital which does not
overlap with similar part of other macro-orbital.  In view of
what has been argued in point (iii) above, the macro-orbital
representation is consistent with MS/SS state of P1 and P2.   

\bigskip
\noindent
(viii) {\it Macro-orbital as an eigenfuntion :}
A macro-orbital ($\xi{(x_{(i)}, X_{(i)})}$) is a
derived form of a wavefunction
which identifies a particle to have two different motions
({\it cf.} Eqn. 15) : the $q$-motion of energy $E(q_{(i)})
= \hbar^2q_{(i)}^2/2m = \hbar^2k_{(i)}^2/8m$ which decides
its quantum size $\lambda_{(i)}/2 = \pi/q_{(i)}$ and
$K$-motion of energy $E(K_{(i)}) = \hbar^2K^2_{(i)}/8m$
which represents a kind of free motion of the particle.
Evidently, this form does not fit, as a solution, with
the form of Schr\"{o}dinger equation (Eqn. 2).  However, 
the corresponding hamiltonian can be rearranged to obtain
Eqn. 2 in a form with which $\xi{(x_{(i)}, X_{(i)})}$ is
compatible and to this end we define
$$h_i = - {\hbar^2\over{2m}}{\partial^2\over
\partial^2_{{\rm x}_i}}, \quad \quad h(i) =
{{h_i + h_{i+1}}\over{2}} \quad \quad {\rm and}
\quad \quad H(2) = h(1) + h(2)   \eqno(16)$$

\noindent
with $i =$ 1 or 2 for a system of $N = 2$, $h_{N+1} = h_1$, 
$h_i$ being the kinetic energy operator of $i$-th particle in
unpaired format of P1 and P2, and $h(i)$ is the same in their
paired format.  We have
$$h(i) = -{\hbar^2\over{8m}}{\partial^2\over
\partial^2_{X_{(i)}}} -{\hbar^2\over{2m}}{\partial^2\over
\partial^2_{x_{(i)}}}  \eqno(17)$$

\noindent
which is such that
$$h(i)\xi{(x_{(i)}, X_{(i)})} = [(E(q)_{(i)} + E(K_{(i)})/2]
\xi{(x_{(i)}, X_{(i)})}.  \eqno(18)$$

\bigskip
\noindent
{\bf 4. N-Particle System and Macro-orbital Representation}

\bigskip
\noindent
(i). {\it Rearrangement of} $H(N)$ : For the consistency of
macro-orbital representation, we use Eqn. 16 to rearrange   
$$H(N) = \sum_{i}^N h_i + \sum_{i>j}A\delta{(x_{ij})}
\eqno(19)$$

\noindent
as
$$H(N) = \sum_{i}^N h(i) + \sum_{i>j}A\delta{(x_{ij})}
\eqno(20)$$

\noindent
or
$$H(N) = \sum_{i}^N\left[
-{\hbar^2\over{8m}}{\partial^2\over \partial^2_{X_{(i)}}} -
{\hbar^2\over{2m}}{\partial^2\over\partial^2_{x_{(i)}}}\right]
+ \sum_{i>j}A\delta{(x_{ij})}. \eqno(21)$$

\bigskip
\noindent
(ii). {\it State Functions} : Using $N$ macro-orbitals for $N$
particles, we construct a state function of the system by
following standard procedure.  Assuming that $\Psi_n$
represents $n$-th state of energy $E_n$, we have
$$\Psi_n = \phi_n(q)\phi_n(K), \eqno(22)$$

\noindent
with 
$$\phi_n(q) = \Pi_{i=1}^N (2/L)^{N/2}\sin{[q_{(i)}x_{(i)}]}, 
\eqno(23)$$

\noindent
and  
$$\phi_n(K)= B\sum_{P}(\pm 1)^P
\Pi_{i=1}^N[\exp{(_PK_{(i)}X_{(i)})}].   \eqno(24)$$

\noindent
Since each particle occupies a cage between its neighbors,
$q_{(i)}$ in Eqn. 23 should be integer multiple of
$\pi/d_{(i)}$ ($d_{(i)} =$ size of the cage);  to a good
approximation, $q_{(i)}$ can also be assumed to be an
integer multiple of $\pi/L$ if $L$ is macroscopically
large.  In Eqn. 24, we have $B = {(N!L^N)}^{-1/2}$ as the
normalization constant, $K_{(i)} = {\rm N}\pi/L$ (with
N = 1, 2, 3, ...) and $\sum_{P}(\pm 1)^P$ as the sum of
$N!$ products of $N$ plane waves obtained by permuting
different $K_{(i)}$ over $N$ particles with $P$ being the
number of performed permutations.  While the use of
$(+1)^P$ or $(-1)^P$ depends on the bosonic or fermionic
nature of particles, the condition $q_{(i)} \ge \pi/d_{(i)}$
follows from the HC nature of particles (bosons and fermions
alike).  Since different particles can, in principle, have
different $q_{(i)}$, their permutation over $N$ particles
renders $N!$ different $\phi_n(q)$ and, therefore, similar
number of $\Psi_n$.  Consequently, the general form of a
state function that should reveal the physics of the system
should be expressed by
$$\Phi_n = (N!)^{-1/2}\sum_{j=1}^{N!}\Psi_n^{(j)}.
\eqno(25)$$

\bigskip
\noindent
(iii). {\it Energy Eigenvalues} : The energy eigenvalue
$E_n$ can be expressed by
$$E_n = {{<\Phi_n|H_o(N) + \sum_{i<j}
A\delta{(x_{ij})}|\Phi_n>} \over{<\Phi_n|\Phi_n>}} = 
{{<\Phi_n|\sum_i^N h(i)|\Phi_n>} \over{<\Phi_n|\Phi_n>}}
\eqno(26)$$

\noindent
which renders
$$E_n = \sum_i^N \left[{\hbar^2K_{(i)}^2 \over 8m} +
{\hbar^2q_{(i)}^2 \over 2m}\right] \eqno(27)$$

\noindent
since $<A\delta{(x_{ij})}>$ vanishes ({\it cf.}, Eqn. 14)
for every pair of HC particles.  One can also conclude
$<A\delta{(x_{ij})}> = o $ by using $N$ body wave
function $\Phi_n$ and the detailed derivations, as used to
conclude similar result for a 3-D $N$ body system [63].

\bigskip
\noindent
(iv). {\it G-state Energy} : While $K_{(i)}$ has zero value
for all bosons in their G-state, its different values
ranging between $0$ and $K_F$ (Fermi wave vector) in the
G-state of a fermionic system follow Fermi-distribution.  To
fix possible values of $q_{(i)}$ for which $E_o$ is minimum,
we note that each particle satisfies $\lambda_{(i)}/2
\le d_{(i)}$ which implies that $\lambda_{(i)}/2$ space
exclusively belongs to a HC particle of momentum $q_{(i)}$;
this agrees with the volume excluded condition envisaged by
Kleban [68] for HC particles in 3-D systems.  Since each
particle in the G-state has lowest energy or largest possible
$\lambda_{(i)}/2$, the net energy of $q$-motion can be
expressed as
$$E_o = \sum_i^N {h^2\over 8md_{(i)}^2} \quad {\rm with}
\quad \sum_i^N d_{(i)} = L \quad ({\rm constant}) \eqno(28)$$

\noindent
by assuming that particles occupy cages of different size
$d_{(i)}$.  Simple algebra reveals that $E_o$ has its
minimum value for $d_{(1)} = d_{(2)} = d_{(3)}...= d_{(N)}
= d$.  Obviously, for a bosonic system  we have
$$E_o^{bose} = N{h^2\over 8md^2} = N\varepsilon_o.
\eqno(29)$$

\noindent
which does not differ from similar result for 3-D bosonic
system [64].  Accounting for an additional contribution
of $K$-motions to the G-state energy of a fermionic system,
we have
$$E_o^{fermi} = N\varepsilon_o +
{1\over 4}\bar{\epsilon_K},  \eqno(30)$$

\noindent
where $\bar{\epsilon_K}$ is the total energy of $K$-motions
at $T=0$ in a system of non-interacting fermions. The factor
(${1\over 4}$) accounts for the fact that each particle of
our system in its macro-orbital representation serves as an
entity of mass $4m$ for its $K$ motion.  For a 3-D system, we
use the well known relations [70] for the Fermi level energy
$E_F \approx h^2/8md^2 = \varepsilon_o$ and
$\bar{\epsilon_K} = {3\over 5}NE_F$ and obtain
$$E_o^{fermi} \approx N\varepsilon_o +
{3\over 20}N\varepsilon_o \approx 1.15N\varepsilon_o.
\eqno(31)$$

\noindent
Howver, for 1-D system, we find $\bar{\epsilon_K} \approx
{1\over 6}NE_F$ and $E_F = {1\over 4}\varepsilon_o$ to get 
$$E_o^{fermi} = N\varepsilon_o + {1\over 24}NE_F
\approx N\varepsilon_o + {1\over 96}N\varepsilon_o
\approx 1.01N\varepsilon_o.  \eqno(32)$$

\noindent
From Eqns. 31 and 32, it is clear that $E_o^{fermi}$ (both
for 3-D system  and 1-D system) is only marginally higher
than $E_o^{bose}$ (Eqn. 29) which indicates that $K$-motions
have very little contribution to the G-state of a fermionic
system ($\approx 15\%$ in case of 3-D system, {\it cf.}
Eqn. 31 and $\approx 1\%$ in case of 1-D system,
{\it cf.} Eqn. 32).

\bigskip
\noindent
(v). {\it Process of Reaching G-State:} In what follows
from the above discussion the free energy ($F$) of the
system can be identified to be a sum of two contributions
$F(K)$ and $F(q)$, respectively, arising from $K-$ and $q-$
motions of particles.  While $F(K)$ refers to a gas of
non-interacting quantum quasi-particles of mass $4m$
constrained to move in one direction only, $F(q)$ could be
equated, to a good approximation, to $N\varepsilon_o$ as
shown for the electron fluid ( a fermionic system ) in a
conductor [66] as well as for liquid $^4He$ ( a bosonic
system ) [64].  As concluded by Mermin and Wagner [71], we
note that $F(K)$ is not expected to show any transition in
case of a 1-D system, while $F(q)= N\varepsilon_o$ does not
depend explicitly on $T$ and $P$.  Consequently, there is
no means to find a relation for a $T$ and $P$ for any
possible transition in our system except by using possible
physical arguments.  In this context, we find that for a
given particle density of the system, $(d - \lambda/2)$
decreases with decreasing $T$ and at certain $T = T_c$, when
$(d - \lambda/2)$ (with $d = L/N$) vanishes at large, $q$
motions get frozen onto zero point motions of $q = q_o =
\pi/d$ and the system moves from a state of $\lambda/2 \le d$
to that of $\lambda/2 = d$.  While $\lambda/2 \le d$
corresponds to $\phi (=2qd) \ge 2\pi$ which represents
randomness of $\phi$ positions and, therefore, a disordered
state in $\phi$-space, $\lambda/2 = d$ defines a state of
$\phi = 2\pi$,- an ordered state in $\phi$-space.  Thus the
system, on its cooling through $T_c$, moves from its
disordered state to ordered state of its particles in the
$\phi-$ space but this change does not represent an onset of
a finite $T$ phase transition.  This is because the change
refers to the relative configuration of particles (relative
position ($x = d$), relative momentum ($k = 2q$) and relative
phase position ($\phi =2qd$)) which remains unaltered with
fall in $T$ from $T = T_c$ to $T = 0$.  Evidently, the said
change at $T = T_c$ leads all the $N$ particles to assume a
configuration of the ground state of their $q-$motions which
naturally have no thermal energy at all $T \le T_c$.  In
other words, the relative configuration of the system is
effectively at $T =0$ and the thermal energy of the system at
all $T \le T_c$ comes entirely from the $K-$motions only.
This clearly means that the said order-disorder of particles
is effectively a $T=0$ change which does not violate
Mermin-Wagner theorem [71] forbidding a transition in 1-D
and 2-D systems at finite $T$.

\bigskip
\noindent
(vi). {\it Merger of $N!$ micro-states} : We note that with
all $q_{(i)} = q_o$, different $\Psi_n^{(j)}$ of $\Phi_n$
(Eqn. 25) become identical and the latter attains the
form of a single $\Psi_n$ (Eqn. 22) which implies that all
the $N!$ microstates merge into one at $T = T_c$ and the
entire system attains a kind of oneness, as envisaged by
Toubes [72]; the system at $T \le T_c$ is therefore
described by
$$\Phi_n(S) = \phi_n^o(q_o)\phi_n^e(K), \eqno(33)$$

\noindent
where $\phi_n^o(q_o)$ and $\phi_n^e(K)$, respectively,
define the relative configuration of the ground state and
the collective excitations of the system.

\bigskip
\noindent
(vii). {\it Quantum Correlation Potential} : The
inter-particle {\it quantum correlation potential} (QCP)
originating from the wave nature of the particles can be
obtained by comparing the partition function (under quantum
limit of the system), $Z_q = \sum_n\exp(-E_n/k_BT)
|\Phi_n(S)|^2$ and its classical equivalent, $Z_c =
\sum_n\exp(-E_n/k_BT)\exp(-U_n/k_BT)$.  Here $\Phi_n(S)$ is
given by Eqn. 33.  This can be justified because $\Phi_n(S)$
is basically nothing but a product of paired wave functions
obtained by simple superposition of plane waves and
our conclusion that two particles in their SMW configuration
satisfy $d \ge \lambda/2$ screens out the HC interaction.
Simplifying $U_n$ one easily finds that the pairwise QCP has
two components. A $U_{ij}^s$, pertaining to the $q$-motion
of particles, controls the $\phi -$position of a particle
and we have
$$U_{ij}^s = -k_BT_o\ln[2\sin^2(k_{ij}x_{ij}/2)], \eqno(34)$$

\noindent
with $k_{ij} = {\rm k}_i - {\rm k}_j$ and $x_{ij} = {\rm x}_i -
{\rm x}_j$ and $T$ replaced by $T_o$ because $T$ equivalent of
$q$ motion energy $\varepsilon_o$ at all $T \le T_c$ is $T_o$.
$U_{ij}^s$ has its minimum value ($-k_BT_o\ln{2}$) at $\phi =
(2n+1)\pi$ and maximum value ($\infty$) occurring periodically
at $\Delta\phi = 2n\pi$ (with $n = 1,2,3,....$).  We note that
$U_{ij}^s$ increases by 
$${1\over 2}{\rm C}(\delta\phi )^2 =
{1\over 4}k_BT_o (\delta\phi )^2  \eqno(35)$$

\noindent
for any small change ($\delta\phi$) in $\phi$ around $\phi =
(2n+1)\pi$ where it has its minimum value ($-k_BT_o\ln{2}$).
This indicates that particles experience a force = -C$\delta
\phi$ (force constant C = ${1\over 2}k_BT_o$) which is
naturally responsible for $\delta\phi = 0$ and an ordered
state in $\phi$-space.  However, since $U_{ij}^s$ is not
the real interaction that may manipulate $d$, the state of
order is achieved by driving all $q$ towards $q_o$.

\smallskip
The second component of QCP, pertains to plane wave $K$
motions, and it can be expressed by
$$U_{ij} = -k_BT\ln{\left[1 \pm \exp{\left(-{2\pi
|X_2 - X_1|^2) \over \lambda_T^2} \right)}\right]},
\eqno(36)$$

\noindent
by following standard procedure [73 and 74] applied to plane
wave motion of non- interacting particles; while $\lambda_T
= h/\sqrt{2\pi (4m)k_BT}$ representing the thermal de Broglie
wave length is associated with $K-$motions (a kind of free
motion where each particle appears to have $4m$ mass),
$+ve(-ve)$ sign stands for a bosonic(fermionic) system.  In
case of a bosonic system, $U_{ij}$ has its minimum
(= $-k_BT\ln 2$) at $|X_2 - X_1| = 0$ implying that $U_{ij}$
facilitates bosons to occupy a common $X$ point.  However,
$U_{ij}$ for a fermionic system has its maximum
(= $\infty$) at $X_2 = X_1$ and, therefore, forbids two
fermions from occupying a common $X$.

\bigskip
\noindent
(viii). {\it Negative Thermal Expansion} : In what follows
from Sections-4(vii), the system on its cooling moves from
its dis-ordered state in $\phi -$space at $T > T_c$ to an
ordered state at $T \le T_c$.  The operational force for
this change lies with : (i) QCP [$U_{ij}^s$, Eqn. 34] which
modulates the $\phi$-positions of particles at $\Delta\phi =
2n\pi$, and (ii) the zero point repulsion ({\it cf.}
Section-3.(vi)) which keeps two particles at $<x> \ge
\lambda/2$.  Consequently, when $\lambda/2$ increases with
falling $T$, each particle pushes its neighbors to make
space for its increased quantum size ($\lambda/2$).  However,
the increase in $\lambda/2$ virtually stops at $T_c$ when
$q$-motions of particles get frozen at $q = q_o$ (or
$\lambda/2 = d$) because the box gets totally occupied by
closely packed wave packets of $N$ particles with
$L = N\lambda/2$.  Evidently the zero-point repulsion
of a particle on its nearest neighbor, at this stage,
renders a force which tries to increase the box size, $L$.
In all practical situations where forces restoring $L$ are
not infinitely strong, a non-zero increase ($+\delta L$)
in $L$ leading to a $-ve$ thermal expansion coefficient
of the system is expected around $T_c$.  The experimental
observation of $-ve$ thermal expansion co-efficient would,
obviously, provide proof for our predictions of : (i) the
freezing of $q$-motions at $q_o$ and (ii) the onset of an
order in the $\phi-$positions of particles.  In this
context, it may be noted that in agreement with our similar
prediction for a 3-D system, liquids $^4He$ and $^3He$
are really found to have -ve thermal expansion coefficient
around 2.17K and 0.55K, respectively [75].

\bigskip
\noindent
(ix). {\it Estimation of $T_c$} : We note that the potential
energy contribution to net energy of the system is zero,
since $<A\delta{(x_{ij})}>$ vanishes for all pairs of
particles.  Even zero-point energy $\varepsilon_o =
h^2/8md^2$, seemingly having potential energy character as
indicated by its $d$ dependence, represents the energy of
residual $q$-motion which implies that the net energy of the
system is kinetic.  Consequently, to a good approximation,
the lower bound of $T_c$ could be equated to $T_{GSE}$ (the
$T$ equivalent of the G-state energy of a particle) with
$T_{GSE} \equiv E^{bose}_o/N$ (Eqn. 29) for the bosonic
system and $T_{GSE} \equiv E_o^{fermi}/N$ (Eqn. 30) for a
fermionic system.   However, to obtain more accurate $T_c$
we need to account for $T_{ex} \equiv E_{ex}$, -the energy of
thermal excitations (representing the correlated $K-$motions
of particles) present in the system at $T_{GSE}$ and this
renders
$$T_c \approx T_{GSE} + T_{ex}.                \eqno(37)$$

\noindent
To estimate $T_{ex}$ for a bosonic system we note that the
freezing of $q$-motions at $q = q_o$, immediately follows an
onset of the condensation of particles in $K=0$ state because
at this stage the system only has $K-$motions that may lose
energy with falling $T$.  Although, as concluded in
Section-3(v), the change at $T_c$ is an effectively $T=0$
transition but one also finds that a quasi 1-D system
realized in different laboratories basically represents a
3-D system where two dimensions are mechanically reduced to
the order of inter-atomic separation by increasing two
corresponding components of momentum of confined particles
and this agrees with the fact that these quasi 1-D systems
do exhibit BEC in a manner identical to 3-D systems.
Evidently, the $K = 0$ condensation in our quasi 1-D
laboratory systems can be identified as the BEC of
non-interacting bosons because particles for their
$K$-motions are represented by plane waves.  Since the
particles for the $K$-motions appear to have $4m$ mass, we
can replace $T_{ex}$ by ${1\over 4}.T_{BEC}$ to  obtain
$$T^{bose}_c = T_o + {1\over 4}T_{BEC}= {h^2\over 8\pi
mk_B}\left[{1\over d^2} + \left({N\over 2.61V}\right)^
{2/3}\right] \eqno(38)$$

\noindent
for a bosonic system, where particles are free to move
within its volume $V$ on a surface of a constant potential.
Similarly, for a system of bosons confined to a harmonic
trap we have
$$T^{bose}_c = T_o + {1\over 4}T_{BEC} \approx
0.55\hbar\omega N^{1/3}/k_B \eqno(39)$$

\noindent
where we use our result $T_o = \hbar\omega N^{1/3}/(\pi k_B)$
[65] as well as $T_{BEC} = \hbar\omega (N/1.202)^{1/3}/k_B$
obtained for such systems by Groot {\it et. al.} [76].

\smallskip
The $T^{bose}_c$ of a hypothetical 1-D system of HC bosons may
be equated to $T_o$ as its lower bound and to $T^{bose}_c$
value given by Eqn. 38/39 as its upper bound since the
$K$-motion energy in such a 1-D system is expected to be lower
than that present in a 3-D system.  This indicates that $T_o$
which represents about $66\%$ of $T^{bose}_c$ in Eqn. 38 and
about $60\%$ in Eqn. 39 would be more close to $T^{bose}_c$ of
a hypothetical 1-D system. 

\smallskip
In order to find $T^{fermi}_c$ (the $T_c$ for a Fermi
system), we similarly use Eqn. 30 to find $T_{GSE} (\equiv
E^{fermi}_o/N)$ as its lower bound and assess $E_{ex}$ at
$T_{GSE}$ to find equivalent $T_{ex}$.  We note that such
$E_{ex}$ should be a small fraction of $E^{fermi}_o$ because
at such a low $T$ not many fermions are in excited state and
$T^{fermi}_c$ can be placed close to $T_{GSE}^+ \approx
T_o^+$ (a $T$ slightly above $T_o$).  It is important to note
that $T_c^{fermi} = T_o^+$ represents a point at which (i)
$q$-motions get frozen at $q = q_o$, (ii) particles define an
ordered state in the phase space and (iii) the system
exhibits $-ve$ {\it thermal expansion coefficient}.  This
should not be confused with a point below which a Fermi system
transforms into a quantum fluid.  As analyzed and discussed in
[65, 66], the superfluid transition point in a Fermi system
falls much below $T_o$.  This is because the fermions always have
non-zero $K$ and according to Pauli exclusion principle two
fermions with equal $q$ have to have unequal $K$ or {\it vice
versa}.  Evidently, when they have equal $K$, they have
unequal $q$ and this possibility does not allow the
relative configuration to stabilize with $\Delta\phi = 2n\pi$
unless the thermal energy falls below their binding energy
which can be there only if the particles have inherent or
induced inter-particle attraction [66].

\bigskip
\noindent
{\bf 5. Concluding Remarks}

\bigskip
\noindent
(i). The paper uses a new scheme to analyze the G-state of a
1-D system of $N$ HC quantum particles.  It concludes that
each particle in the system should better be described by a
macro-orbital ({\it a kind of pair wave form}) since each
particle represents a part of the pair of particles ({\it
the basic unit of the system}) having equal and opposite
momenta ($q$, -$q$) at their CM which moves with momentum $K$
in the laboratory frame.  This agrees with the fact that
particles in the system interact through a two body potential
and their individual momenta (${\rm k}_i$ with $i = 1, 2, 3,
...N$) do not define good quantum numbers as soon as the
particles have their wave mechanical superposition.  Moreover
the fact, that a particle in its macro-orbital representation
has two motions [{\it viz.}, the $q-$ and $K-$ motions]
of the pair, -it represents, is consistent with Eqn. 6.

\bigskip
\noindent
(ii). While all particles in the G-state of both systems
(bosonic as well as fermionic) retain $q$-motions of
identically equal $q = q_o = \pi/d$, they all have $K=0$
for bosonic system and follow Fermi-distribution to occupy
different $K-$states ($K = n\pi/L$ with $n$ = 1,2,3, ...)
with $K$ ranging from $\pi/L \approx 0$ to $K_F$ in case of a
fermionic system.

\bigskip
\noindent
(iii). Particles in the G-state of both systems define a
close packed arrangement of their representative wave packets
(each of size $\lambda/2 = d$) with inter-particle phase
separation $\Delta\phi = 2n\pi$ ($n$ = 1,2,3, ...), nearest
neighbor distance $<x> = d$ and this packing does not allow
them to have
relative motion.  Consequently, the G-state is a state of
collisionless motion which can be identified as the residual
zero-point $q$-motion.  It differs from the higher energy
states where particles can be perceived to have collisions
because their effective size ($\lambda/2$) is smaller than
the average space ($d$) available to each particle.

\bigskip
\noindent
(iv). While a bosonic system is expected to exhibit $-ve$
thermal expansion coefficient at $T \approx T_c^{bose}$, a
fermionic system should show such effect around $T_c^{fermi}
\approx T_o$ ({\it cf.} Section-4(viii)).  This agrees with
similar conclusion for a single particle in 1-D box [61] and
two HC particles in similar box [58].  Our scheme applied to
3-D systems [63-66] also predicts similar expansion whose
accuracy is established by experimentally observed expansion
of systems like liquids $^4He$ and $^3He$ [75].

\bigskip
\noindent
(v). The G-state configuration of our system ({\it cf.}
point (ii), above) does not assume stability against any
perturbation ({\it viz.}, the flow of the system) because
the particles in this configuration are in a state of
persisting zero-point repulsion and the necessary
inter-particle attraction to counter this repulsion is
absent.  If such an attraction exists and the system
remains fluid even at $T \le T_c$, then the system can, in
principle, exhibit phenomenon like superfluidity below
certain $T$ (say $T_{\lambda}$); our analysis of 3-D systems
such as liquid $^4He$ [64] and liquid $^3He$ [65, 66]
concludes that $T_{\lambda}$ for a bosonic
system falls close to $T_o$ but for a fermionic system it is
expected to be orders of magnitude lower than $T_o$ because
$K-$motion energy retained by fermions due to Pauli exclusion
is good enough to de-stabilise the G-state configuration
unless $k_BT$ falls below per particle binding energy. In our
forthcoming paper, we plan to study different aspects of
our 1-D system when weak interparticle attraction is added
as a perturbation.  However, our detailed study of a 3-D
bosonic system [64] and a brief qualitative analysis of widely
different many body systems (including low dimensional systems
and fermionic systems) [65] clearly reveals that a 1-D system
realised in a laboratory must have basic properties such as
superfluidity and related behavior of 3-D systems if weak
inter-particle attraction is present.  As shown in Section-4(v)
the predicted order-disorder transformation of our 1-D
systems is consistent with Mermin Wagner theorem [71].  In
this context, it may be emphasized that the transformation
is basically a quantum transition expected to occur at
$T = 0$; however, it occurs at a non-zero $T$ for the
proximity of zero-point $q-$motions ($q = q_o$) with
$K-$motions which solely represent the thermal motions of
particles at $T \le T_o$.  

\bigskip
\noindent
(vi). The formation of a SMW from the superposition of two
plane waves of two HC particles is as natural as the phenomena
of interference and diffraction of particles such as strongly
interacting electrons, neutrons, helium atoms, {\it etc.} [77,
78].  Since the nature of interference and diffraction patterns
of these strongly interacting particles does not differ from
the nature of such patterns for non- interacting photons, it
is evident that only wave nature (not the inter-particle
interactions) modulates the relative phase positions
of particles in their wave superposition.  Evidently, the fact
that these experiments support the formation of a SMW can not
be doubted.

\bigskip
\noindent
(vii). We understand that, in principle, two particles
described by plane waves should have their superposition
independent of their $d$ and $\lambda$.  However the
experimental fact, that wave nature dominates the behavior
of a many body system only when $\lambda \ge d$ [59, 69],
indicates that {\it effective wave superposition} of two
particles becomes possible only when $\lambda \ge d$ and
this could be so because no particle in nature manifests a
real plane wave.  Since the important consequences of wave
superposition ({\it cf.} Section-3.0) such as zero point
repulsion leading to $<x> \ge \lambda/2$, inter-particle
$\phi-$correlation which arranges particles at $\Delta\phi
= 2n\pi$ and renders coherence of particle motion,
{\it etc.} (incorporated in the present analysis) are not
used in [1-6], our results are expected to differ from
those of [1-6].  Guided by this observation and the
experimental support to the formation of a SMW (Point-vi,
above), we may highlight the said differences by comparing
our results with those of [4] for $\gamma = \infty$ case.
If our inference for allowed $q = q_o = \pi/d$ and $K =
\pi/L \approx 0$ for the G-state of a bosonic system is
used to determine the corresponding momenta through
${\rm k}_i =  q + K/2$ and ${\rm k}_j = - q + K/2$ of
two particles in superposition, each boson should either
have $\pi/d$ or $-\pi/d$ momentum and we find that these
values differ from possible ${\rm k} = \pm s\pi/L$ (with $s$ =
1,2,3, ...) considered in [4].  Similar estimates of ${\rm k}$
for a fermionic system using $q = q_o = \pi/d$ and $K$ ranging
between $\pi/L \approx 0$ and $K_F = N\pi/2L = \pi/2d$ render a
band of allowed ${\rm k}$ ranging from $-q+K_F/2 = 3\pi/4d$ to
$q+K_F/2 = 5\pi/4d$ (with least difference of $\pi/4L$) which
do not match with ${\rm k} = \pm 2s\pi/L$ considered in [6].
Note that these allowed ${\rm k}$ have been determined just for
their comparison with those of [1-6], otherwise our analysis
reveals that ${\rm k}_i$ and ${\rm k}_j$ do not represent good
quantum numbers for the two particles in a state of wave
superposition.  Similarly, per particle G-state energy
$\varepsilon_o = h^2/8md^2$ obtained by us for a bosonic
system is three times higher than $(1/3)\varepsilon_o$
concluded in [4].  It may be noted that our results reveal a
well defined picture of the G-state of 1-D systems.
Accordingly, the system in its G-state is a close
packed arrangement of the wave packets of its particles with
$<x> = d$, $q = q_o$ and $\Delta\phi = 2n\pi$ and this
fact has been revealed for the first time.  One may
find a number of other points where our results differ from
those of [1-6].  However, since an effective wave
superposition of particles does not exist in higher energy
states of $\lambda < d$, such particles can equally well be
described by plane waves as considered in [1-6].  Evidently,
such studies seem to provide reasonably accurate
understanding of higher energy states (or high $T$ phase) of
the system and the present study, revealing the low energy
states (or low $T$ phase), can be identified to supplement
the results of [1-6] in providing a complete understanding
of a 1-D system.  In view of an equivalence in the dynamical
behavior of two HC particles interacting through $V_{HC}$
and $A\delta{(x)}$, particularly, for low energy states of
$\lambda/2 \ge \sigma$, it can be stated that our analysis
also applies to particles of finite $\sigma$ satisfying
$\lambda/2 \ge \sigma$.  Finally, as an important inference
of this study, the true picture of the low energy states
of our system can be revealed if its theory incorporates
the impacts of the wave superposition and zero-point repulsion
({\it obvious effects of wave particle duality}), {\it viz.,}
$<x> \ge \lambda/2$ $q \ge \pi/d$ and $\Delta\phi \ge 2n\pi$
on the relative configuration of two particles.

\bigskip
\noindent
{\it Acknowledgement} : The author is thankful to Dr. P. K.
Patra for useful discussion.

\end{document}